\documentclass[10pt]{iopart}
\usepackage{graphicx}
\begin{document}

\title[]{On the rotating wave approximation in the adiabatic limit}

\author{Jonas Larson}

\address{Department of Physics, Stockholm University, Albanova University Center, Se-106 91 Stockholm, Sweden}
\ead{jolarson@fysik.su.se}
\begin{abstract}
I revisit a longstanding question in quantum optics; When is the rotating wave approximation justified? In terms of the Jaynes-Cummings and Rabi models I demonstrate that the approximation in general breaks down in the adiabatic limit regardless of system parameters. This is explicitly shown by comparing Berry phases of the two models, where it is found that this geometrical phase is strictly zero in the Rabi model contrary to the non-trivial Berry phase of the Jaynes-Cummings model. The source of this surprising result is traced back to different topologies in the two models.
\end{abstract}

%\maketitle

\section{Introduction}
The rotating wave approximation (RWA) is taught in any graduate course on quantum optics~\cite{scully}. One of the more familiar examples were the RWA has proven very efficient is for the Jaynes-Cummings (JC) model describing the interaction between a single two-level atom and one quantized resonator mode~\cite{scully,jc1}. For moderate field amplitudes and quasi-resonant interaction between the field and atomic transition frequencies, the approximation is taken to be justified whenever the atom-field coupling is small in comparison to the mode frequency. This has been verified by numerous cavity QED experiments~\cite{haroche}, where the JC model correctly predicts the measured quantities.

It is not until very recently that the applicability of the RWA in the adiabatic limit was questioned~\cite{jonasBerry}. We know that for very long evolution times, small perturbations can accumulate and greatly affect the results. Of particular interest is when the perturbation alters the symmetry of Hamiltonian, or even its topology. In this work I show how this is exactly the case for the JC model. At a mean-field level, the adiabatic energy surfaces in phase space change topology when the RWA is applied. In particular, the surfaces of the JC model possesses a {\it conical intersection} (CI) absent for the Rabi model (JC model including the counter rotating terms). Such a CI is the origin of non-trivial Berry phases~\cite{baer}, and as a result one concludes that the commonly considered Berry phase must strictly vanish in the Rabi model while it is non-trivial for the JC model. Employing numerical diagonalization, going beyond the above mean-field arguments, it is also demonstrated that the so called ``vacuum induced Berry phase'' vanishes.     

\section{Adiabatic concepts}
Adiabaticity has many appearances in quantum mechanics. The original idea dating back to 1928~\cite{adtheo} and given in most advanced text books on quantum mechanics considers an explicitly time-dependent Hamiltonian $\hat{H}(t)$, and the {\it adiabatic theorem} states that a system prepared in an instantaneous eigenstate of $\hat{H}(t)$ remains in this state provided the change of $\hat{H}(t)$ is ``slow enough'' and there is an energy gap to nearby eigenenergies. While the adiabatic theorem deals with strictly time-dependent systems, the notion of adiabaticity can be efficiently translated to other situations where the time-dependence is replaced by an intrinsic evolution. The system evolution can, in many cases, be characterized by various time-scales. Whenever these scales are very different, it become legitimate to perform an adiabatic separation or elimination of the fast degrees-of-freedom. This is the basis of the {\it Born-Oppenheimer approximation}~\cite{baer,bo}. 

Consider a system with two well separated characteristic time-scales. Letting $R$ and $r$ denote dynamical variables for the slow and fast degrees-of-freedom respectively, we write the Hamiltonian $\hat{H}(R,r)$. Within the BOA we assume the fast degrees-of-freedom to instantaneously follow the motion of the slow degrees-of-freedom. Thus, freezing $R$ we can solve the eigenvalue problem $\hat{H}(R,r)|\Phi(R,r)\rangle=V_{ad}(R)|\Phi(R,r)\rangle$, where $|\Phi(R,r)\rangle$ is the {\it adiabatic eigenstate} and $V_{ad}(R)$ the {\it adiabatic potentials}. The adiabatic eigenstates (labeled them with some quantum number(s) $n$) form a complete basis for the full system state $|\Psi(R,r)\rangle$. The BOA consists in neglecting the ("non-adiabatic'') couplings $A(R,r)$ between these adiabatic eigenstates, and the corresponding coefficients $\phi_n(R)$ in an expansion of $|\Psi(R,r)\rangle$ are given by the eigenvalue problem $\hat{H}(R)\phi(R)=\left[\hat{T}+V_{ad}(R)\right]\phi_n(R)=E\phi_n(R)$, where we symbolize the ``kinetic'' energy operator for the slow degrees-of-freedom with $\hat{T}$.

For this work, where we study a spin 1/2 particle coupled to a single boson mode, the slow degree-of-freedom is the boson mode representing an electromagnetic resonator mode and the internal two-level structure of the particle is the fast degree-of-freedom. Typically, the characteristic instantaneous time-scales depend on the full state $|\Psi(R,r)\rangle$ and if there is no clear separation in the time-scales the BOA breaks down, which become evident in the fact that the non-adiabatic couplings $A(R,r)$ are no longer negligible. Note, however, that the absolute size of $|A(R,r)|$ does not serve as a measure of adiabaticity, but it must be put in relation to the actual dynamics~\cite{jonasad}. $A(R,r)$ is the so-called {\it Mead-Berry connection} and has the properties of a gauge potential and is in general on matrix form~\cite{baer}. The gauge freedom lies in the ambiguity of an overall ($R$-dependent) phase of the adiabatic states $|\Phi(R,r)\rangle$.

\section{Jaynes-Cummings vs. Rabi}
\subsection{Breakdown of the RWA in the ultrastrong coupling regime}
\begin{figure}[h]
\centerline{\includegraphics[width=8cm]{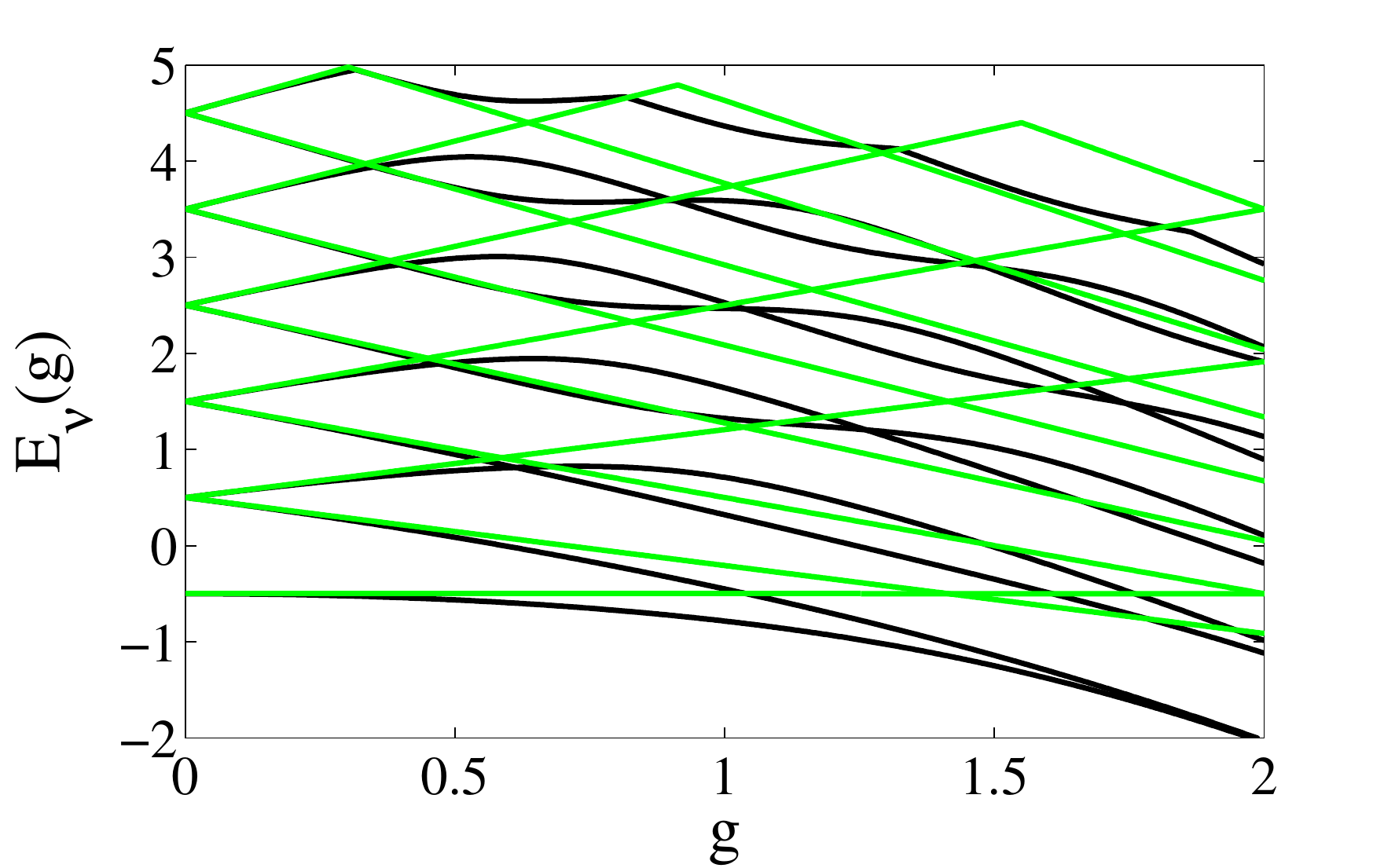}}
\caption{The eleven first eigenvalues $E_\nu(g)$ ($\nu=1,\,2,...,11$) for the Rabi (black) and JC (green) models. The frequencies are $\omega=\Omega=1$. } \label{fig1}
\end{figure}

For the last three decades, the JC model has served as a workhorse in cavity QED~\cite{jc1}, explaining and predicting many of the cavity QED experiments to date~\cite{haroche}. With the development of circuit QED~\cite{wallraff1}, the validity of the RWA applied in the derivation of the JC model has been thoroughly explored, see for example~\cite{circuitRWA}. The JC model is derived from the Rabi Hamiltonian ($\hbar=1$)
\begin{equation}\label{rabiham}
\hat{H}_R=\omega\hat{a}^\dagger\hat{a}+\frac{\Omega}{2}\hat{\sigma}_z+g\sqrt{2}\left(\hat{a}^\dagger+\hat{a}\right)\hat{\sigma}_x
\end{equation}
by neglecting the counter rotating terms $\hat{a}^\dagger\hat{\sigma}^+$ and $\hat{a}\hat{\sigma}^-$ corresponding to virtual light-matter excitations, i.e.
\begin{equation}\label{jcham}
\hat{H}_{JC}=\omega\hat{a}^\dagger\hat{a}+\frac{\Omega}{2}\hat{\sigma}_z+g\sqrt{2}\left(\hat{a}^\dagger\hat{\sigma}^-+\hat{\sigma}^+\hat{a}\right).
\end{equation}
Here $\omega$ and $\Omega$ are the respective resonator and qubit (atom) transition frequencies, $g$ the effective qubit-light coupling, $\hat{a}^\dagger$ ($\hat{a}$) is the photon creation (annihilation) operator, and the $\sigma$-operators are the standard Pauli matrices acting on the internal qubit states $|1\rangle$ and $|2\rangle$. In this setting, the justification of the RWA is typically given by $\omega\gg g$~\cite{jonas1}. While cavity QED experiments operate in the regime $10^{-7}\leq q/\omega\leq10^{-5}$~\cite{haroche}, the first generation circuit QED realizations reached $g/\omega\sim10^{-2}$~\cite{wallraff2}, and more recently $g/\omega\sim0.1$ has been achieved and as a consequence the Bloch-Siegert shift measured~\cite{mooij}. For $g\gg\kappa,\,\Gamma$ where $\kappa$ is the characteristic resonator decay rate and $\Gamma$ the decay of the qubit, the short time evolution is predominantly coherent and the system is said to work in the {\it strong coupling regime}. The works of ref.~\cite{mooij} operate instead in the {\it ultrastrong coupling regime} in which $g/\omega>0.1$. The eleven lowest eigenenergies for the Rabi and JC Hamiltonians as a function of the coupling strength are shown in figure~\ref{fig1}. Throughout this work we use $\omega=1$, i.e. frequencies (energies) are scaled in terms of the photon energy. In agreement with the above mentioned experiment, at around $g\sim0.1$ the Bloch-Siegert shift becomes clearly visible with a relative error $\sim0.1$ $\%$. Further, at $g=\sqrt{2}$ the JC spectrum shows a crossing and the eigenstate $|0,1\rangle$ (vacuum and deexcited qubit) is no longer the ground state. For the same coupling, when including $N$ identical qubits, hence considering the Dicke model~\cite{dicke}, the Rabi model possesses a thermal~\cite{dickept} as well as quantum phase transition~\cite{dickept2} from a normal to a superradiant phase.     

The conservation of excitations $\hat{N}=\hat{n}+\frac{1}{2}\hat{\sigma}_z$ ($\hat{n}=\hat{a}^\dagger\hat{a}$) of the JC model results in a continuous $U(1)$ symmetry and thereby integrability. Together with the ``empty'' eigenstate $|0,1\rangle$, the eigenstates are the {\it dressed states},
\begin{equation}\label{jceig}
\begin{array}{c}
|\Phi_n^+\rangle=\cos(\theta)|n-1,2\rangle+\sin(\theta)|n,1\rangle,\\ \\
|\Phi_n^-\rangle=\cos(\theta)|n-1,2\rangle+\sin(\theta)|n,1\rangle,
\end{array}
\end{equation}
where $\tan(2\theta)=2g\sqrt{2n}/\Delta$ with $n$ a positive integer and $\Delta=\Omega-\omega$ the qubit-field detuning. The counter rotating terms break the $U(1)$ symmetry and the Rabi model has instead a discrete $Z_2$ parity symmetry. It was not until recently that it was demonstrated that also the Rabi model is (quantum) integrable~\cite{braak}. The analytic solutions, however, are not on simple closed form; both the spectrum and the eigenstates are given as sums of complicated functions. Thus, the search for simple approximate expressions for the solutions of the Rabi model has continued, see among others~\cite{rwaapp}. 

\subsection{Mean-field Born-Oppenheimer approach}  
\begin{figure}[h]
\centerline{\includegraphics[width=8cm]{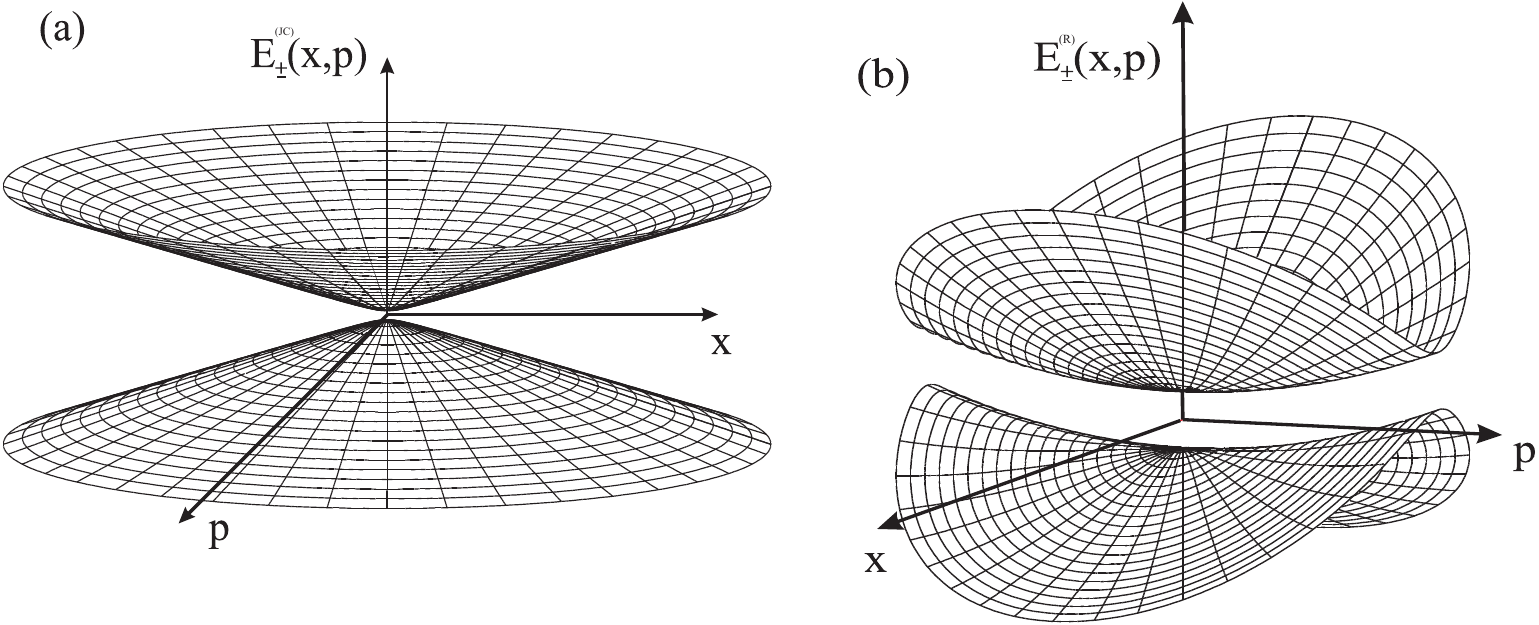}}
\caption{The semiclassical energy surfaces for the JC (a) and Rabi (b) models. The JC model shows a CI absent in the Rabi model.} \label{fig2}
\end{figure}

Let us now employ the adiabatic ideas of the previous section to the JC and Rabi models. We have the degree-of-freedom for the boson mode and that of the qubit. As already pointed out, the validity of the BOA will depend on the actual state we consider. We begin by rewriting the Hamiltonians in a quadrature representation
\begin{equation}\label{jcham2}
\hat{H}_{JC}=\frac{\Delta}{2}\hat{\sigma}_z+g\left(\hat{x}\hat{\sigma}_x+\hat{p}\hat{\sigma}_y\right)
\end{equation}
and
\begin{equation}\label{rabiham2}
\hat{H}_{R}=\omega\left(\frac{\hat{p}^2}{2}+\frac{\hat{x}^2}{2}\right)+\frac{\Omega}{2}\hat{\sigma}_z+2g\hat{x}\hat{\sigma}_x.
\end{equation}
Here we give the JC Hamiltonian in the interaction picture, and the quadratures are
\begin{equation}
\hat{x}=\frac{1}{\sqrt{2}}\left(\hat{a}+\hat{a}^\dagger\right),\hspace{2cm}\hat{p}=\frac{i}{\sqrt{2}}\left(\hat{a}-\hat{a}^\dagger\right).
\end{equation}
In the terminology of the previous section, $R$ represents $\hat{x}$ and $\hat{p}$ while $r$ serves as the $\hat{\sigma}_\alpha$ ($\alpha=x,\,y,\,z$). Applying the BOA, we diagonalize (fast evolving) spin degrees-of-freedom under the assumption of commutability of $\hat{x}$ and $\hat{p}$. Within this approximation we identify the resulting ``diagonal'' Hamiltonian as semiclassical energy surfaces $E_\pm(x,p)$, explicitly given by
\begin{equation}
\begin{array}{c}
E_\pm^{(JC)}(x,p)=\pm\displaystyle{\sqrt{\frac{\Delta^2}{4}+g^2\left(x^2+p^2\right)}=\pm\sqrt{\frac{\Delta^2}{4}+g^2r^2}},\\ \\
E_\pm^{(R)}=\displaystyle{\omega\left(\frac{p^2}{2}+\frac{x^2}{2}\right)\pm\sqrt{\frac{\Omega^2}{4}+4g^2x^2}}
\end{array}
\end{equation}
and displayed in figur~\ref{fig2}.  

The semiclassical energy surfaces of figure~\ref{fig2} make clear that the presence or absence of counter rotating terms change the general structure. For the JC model there is a CI degeneracy at the origin for vanishing detuning $\Delta=0$. In the Rabi model there is not a single point degeneracy but a seam along the $p$-axis. The energy gap $\delta E=E_+(x,p)-E_-(x,p)$ sets the characteristic time-scale for the spin, and it follows that in general large photon field state implies justification of the BOA, i.e. for $\bar{n}\equiv p^2/2+x^2/2\gg0$. The exception occurs for $x\approx0$ in the Rabi model where the gap is simply $\delta E=\Omega$.

\section{Adiabaticity and Berry phases in cavity/circuit QED}
Symmetry, topology and geometry play important roles in physics. For example, two Hamiltonians can have nearby identical spectrum, but different symmetries or topologies which may have far-reaching consequences. In recent years these aspects have seen especially renewed interests, for example in topological quantum computing~\cite{topcomp} and topological insulators~\cite{topins}. Other known examples worth mentioning and related to this work include:

\begin{enumerate}
\item {\it Geometric Berry phases.} When some state vector $|\Psi(R)\rangle$, being an instantaneous non-degenerate gaped eigenstate of the Hamiltonian, is taken adiabatically around a closed loop $C$ in parameter space $R$, it acquires an overall additional phase $\gamma_C$ to the dynamical one~\cite{berry}. Hence, the parameters $R$ are indirectly time-dependent. This phase has a geometric origin and depends on the loop $C$, but not on the actual time it takes to enclose the loop. Explicitly it reads~\cite{berry}
\begin{equation}\label{berryex}
\gamma_C=\oint_CdR\cdot A(R)\equiv i\oint_CdR\cdot\langle\Psi(R)|\nabla_R|\Psi(R)\rangle,
\end{equation}
where we have reintroduced the Mead-Berry connection $A(R)$~\cite{baer}. As for the Mead-Berry connection for the Born-Oppenheimer treatment, $A(R)$ is again comprising the non-adiabatic couplings. The Berry phase (\ref{berryex}) is defined in the strict adiabatic limit, where we must have that the non-adiabatic couplings vanish. Nevertheless, even though the non-adiabatic couplings are zero we may have $\gamma_C\neq0$ (or any multiple of $2\pi$). This reflects the geometric property of the Berry phase; $\gamma_C$ must depend on the geometry of the energy surfaces in parameter space.

\item {\it Reduction of symmetry.} We already touched upon the fact that the continuous $U(1)$ symmetry of the JC model is reduced to a discrete $Z_2$ symmetry when the counter rotating terms are included. This implies that the corresponding classical models (obtained in a mean-field coherent state ansatz) are either integrable (JC) or non-integrable (Rabi). As a consequence, the JC model does not show classical chaos while the Rabi one does~\cite{rabichaos}. 

\item {\it Non-commutability between semiclassical and adiabatic limits.} The classical limit of quantum mechanics is often taken as $\hbar\rightarrow0$. The adiabatic limit is assumed to be that for which any characteristic time-scale $T$ goes to infinity, $\nu\equiv T^{-1}\rightarrow0$. It has been argued that these two limits are equivalent~\cite{Hwang}. This was later proved wrong by considering a counter example in terms of a Landau-Zener problem~\cite{berrysem}. More precisely, the order in letting $\hbar$ and $\nu$ approach zero may render qualitatively different results. 
In a later reference~\cite{wusem}, where a non-linear Landau-Zener model was analyzed, it was demonstrated that the non-cummatibility between the two limits appeared when the instantaneous eigenergies change topology. 

\item {\it Secular approximation.} Utilizing several approximations is the standard procedure when deriving master equation describing the reduced dynamics of a system weakly coupled to an environment. The {\it Born approximation} neglects correlations between the system and the bath, the {\it Markovian approximation} disregards memory-effects of the bath, and finally the {\it secular approximation} is a version of the RWA~\cite{ct}. The application of the secular approximation casts the master equation into a {\it Lindblad form} known to preserve many physical properties~\cite{lidar}. Despite this appealing form, it has been shown that the secular approximation can result in unphysical results~\cite{mottonen}.

\end{enumerate}

Having argued that the topology of the energy surfaces in parameter space can affect system properties greatly, we return to the JC and Rabi models. Already figure~\ref{fig2} signals a qualitative difference between the two models. The semiclassical energy surfaces of the JC model possesses a CI. It has long been known, both in molecular/chemical physics~\cite{baer,lh} and in condensed matter theories~\cite{graph}, that CI's render the system dynamics, and it is especially recognized that encircling a CI adiabatically results in a non-trivial geometric Berry phase. At the same time, it is known that without CI's there are only trivial Berry phases~\cite{baer}. It is thereby expected that the JC and Rabi models give different Berry phases when encircling the origin $(x,p)=(0,0)$. 

Our two models are time-independent and to compass the origin we apply the unitary operator $\hat{U}(\phi)=\exp\left[-i\hat{n}\phi\right]$ to our Hamiltonians~\cite{vlatko}, and think of $\phi$ as an external controllable parameter that is varied adiabatically $\phi:\,0\rightarrow2\pi$. Experimentally, such a change can be accomplished via an external driving field with adjustable phase. The effect of $\hat{U}(\phi)$ is to rotate the state in phase space, i.e. bringing about the encircling of the origin. 

Employing the BOA we can directly calculate the resulting Berry phase for the JC model~\cite{jonas2}  
\begin{equation}\label{berryjc}
\gamma_\pm^{(CI)}=\int_0^{2\pi}d\phi\,\langle\Phi_\pm(\phi)|\frac{d}{d\phi}|\Phi_\pm(\phi)\rangle=\pm\pi\left(1-\frac{\Delta/2}{\sqrt{\frac{\Delta^2}{4}+g^2\rho^2}}\right),
\end{equation}
where $|\Phi_\pm(\phi)\rangle$ are the two adiabatic eigenstates of the JC model, and $\rho$ is related to the field amplitude as $\rho^2/2=p^2/2+x^2/2=n+1/2$. The above result relies an the BOA where we neglect corrections from the non-commutability between $\hat{x}$ and $\hat{p}$. If we compare to the exact result using the eigenstates (\ref{jceig}) one finds an 1/2 error~\cite{vlatko}:
\begin{equation}
\gamma_\pm^{(JC)}=\pm\pi\left(1-\frac{\Delta/2}{\sqrt{\frac{\Delta^2}{4}+g^2(n+1)}}\right).
\end{equation}
In the large field limit, where the BOA becomes exact, the two results agree. We may rephrase the mean-field Born-Oppenheimer analysis by directly study the unitary transformed Hamiltonian 
\begin{equation}
\begin{array}{lll}
\hat{H}'_{JC}(\phi) & = & \hat{U}(\phi)\hat{H}_{JC}\hat{U}^{-1}(\phi)\\ \\
& = & \!\!\frac{\Delta}{2}\hat{\sigma}_z\!+\!\frac{g}{\sqrt{2}}\left[\left(\cos\phi\,\hat{x}+\sin\phi\,\hat{p}\right)\hat{\sigma}_x\!+\!\left(\sin\phi\,\hat{x}-\cos\phi\,\hat{p}\right)\hat{\sigma}_y\right].
\end{array}
\end{equation}
Introducing an effective magnetic field $\hat{{\bf B}}(\phi)=(B_x,B_y,B_z)=\left(\cos\phi\,\hat{x}+\sin\phi\,\hat{p},\sin\phi\,\hat{x}-\cos\phi\,\hat{p},\Delta/2\right)$ we can write $\hat{H}'_{JC}(\phi)=\hat{{\bf B}}(\phi)\cdot\hat{{\bf \sigma}}$ mimicking a spin 1/2 particle in a magnetic field. Since all three field components $B_x$, $B_y$ and $B_z$ are in general non-zero there is no way the adiabatic eigenstates $|\Phi_\pm(\varphi)\rangle$ can become strictly real-valued and $\gamma_{JC}$ will typically be non-trivial.

Turning now to the Rabi model we see that the transformed Hamiltonian reads
\begin{equation}
\hat{H}'_R=\omega\left(\frac{\hat{p}^2}{2}+\frac{\hat{x}^2}{2}\right)+\frac{\Omega}{2}\hat{\sigma}_z+2g\left(\cos\phi\,\hat{x}-\sin\phi\,\hat{p}\right)\hat{\sigma}_x.
\end{equation}
The effective $y$-component magnetic field $B_y=0$ implying that there is a gauge freedom to choose the adiabatic eigenstates real-valued with a vanishing Berry phase as a result. As we already pointed out, this result follows also directly from the semiclassical energy surfaces of figure \ref{fig2} since the Rabi surfaces $E_\pm^{(R)}(x,p)$ lack any CI's~\cite{baer}. 

\begin{figure}[h]
\centerline{\includegraphics[width=8cm]{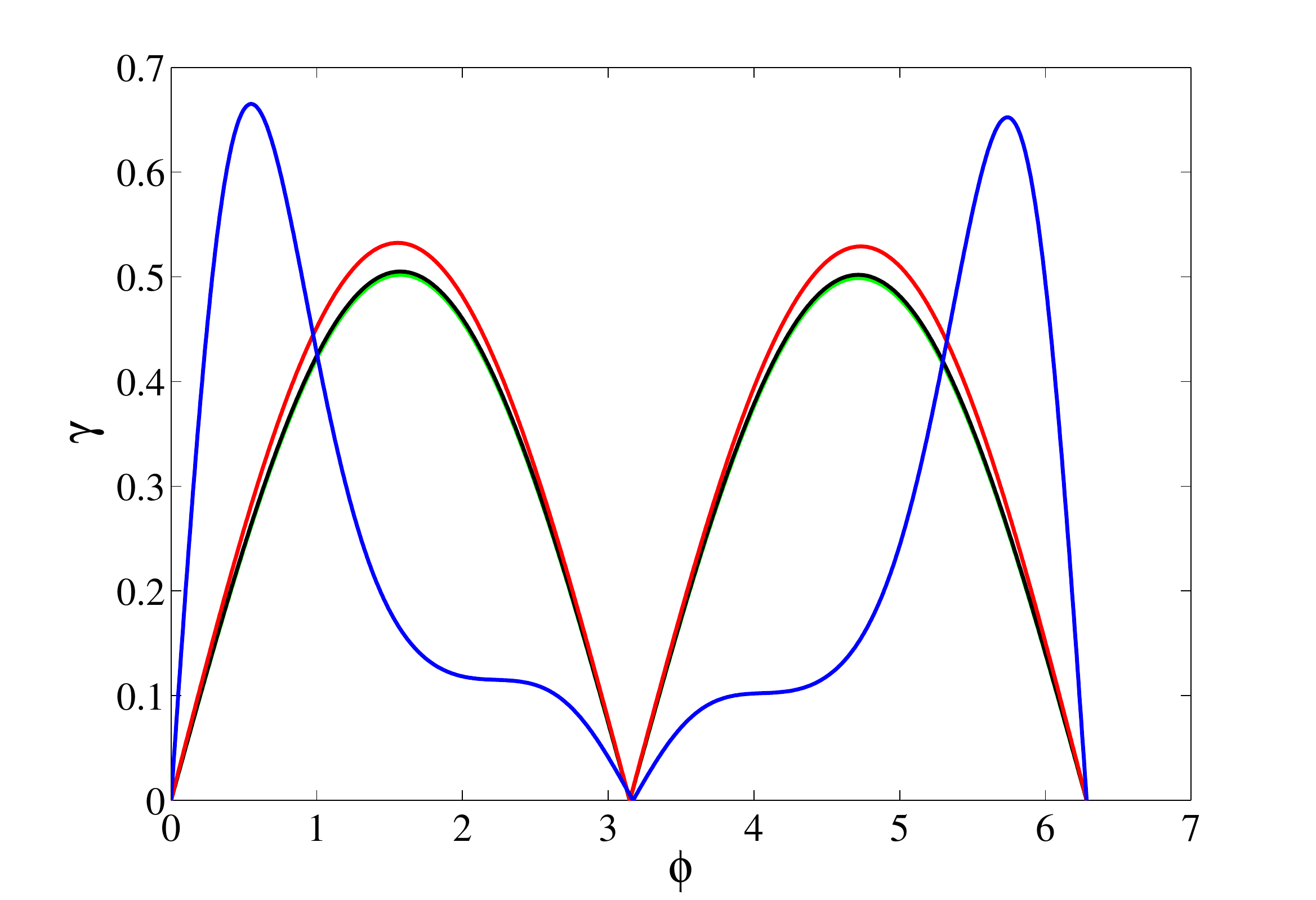}}
\caption{Numerically calculated Berry phase for the first excited state of the Rabi model. The lines correspond to different coupling strengths, $g=0.001$ (green), $g=0.01$ (black), $g=0.1$ (red) and $g=1$ (blue). Note that independent of $g$, the Berry phase always vanishes after closing the loop $C$ at $\phi=2\pi$. } \label{fig3}
\end{figure}

The absence of a Berry phase in the Rabi model as declared above relies on mean-field approximations, and it is not evident that these findings hold true also for small field amplitudes. Nevertheless, the change in topology in the energy surfaces between the two models suggests that also in the weak field limit the Berry phase should disappear. We verify this statement by numerically calculating the Berry phase for the Rabi model. The eigenstates of the model are obtained by diagonalizing the Hamiltonian $\hat{H}'_R(\phi)$ expressed in the bare basis $\{|n,1\rangle,\,|n,2\rangle\}$ where $n$ is a non-negative integer truncated at some large value ($\approx500$). The Berry phase (\ref{berryex}), (\ref{berryjc}) is then calculated by numerically evaluating the integral for $\phi:\,0\rightarrow2\pi$. We consider the first excited state since the groundstate has a vanishing Berry phase also for the JC model~\cite{vlatko}. In the JC model, this state can be thought of as initially the atom is excited while the cavity is in vacuum. The results are displayed in figure~\ref{fig3} showing how the Berry phase varies as $\phi:\,0\rightarrow2\pi$ for different couplings $g$. What is clear is that regardless of $g$, the phase is strictly zero when closing the loop ($\phi=2\pi$). This demonstrates how ``vacuum induced Berry phases'' do not exist when including the counter rotating terms. We may note that $g=0.001$ is in between cavity QED and typical circuit QED experiments, $g=0.01$ is in 
agreement with the circuit QED experiments~\cite{wallraff2}, while $g=0.1$ are in the ultrastrong coupling regime recently demonstrated in~\cite{mooij}.  
 
\section{Concluding remarks}
The results presented in this work can be taken as being in conflict with earlier studies~\cite{vlatko,jcBerry}. However, all of them start from a rotating wave Hamiltonian and as demonstrated here, it is then not surprising that the Berry phase is non-trivial. Interestingly, long before the work by Fuentes-Guridi {\it et al.} the vacuum Berry phase of the JC model had been considered in reference~\cite{klimov} with contradicting results, i.e. they claim that this phase vanishes also in the JC model. It should be noted though, that the employed WKB method of~\cite{klimov} relies on semiclassical arguments. Despite the fact that the Berry phase seems to disappear in the Rabi model, I am not saying that irrelevant Berry phases are not achievable in cavity/circuit QED systems. Other system configurations can indeed generate topologically non-trivial energy surfaces, for example in bimodel set-ups~\cite{jonas3}.

Another relevant aspect concerns the accuracy of the JC or Rabi models to describe certain physical systems. Already performing the single mode and two-level approximations utilize sort of adiabatic elimination schemes, and it is not clear whether such terms can as well influence the topology and thereby alter results in the adiabatic limit. It is generally believed that the Rabi model presents a more precise description of current cavity and circuit QED experiments than that of the JC model. This seems as a trivial statement, but the issue is indeed more subtle than believed~\cite{beige}.

Maybe the most striking outcome of this work relates to the experiments that claim to have measured the Berry phase in circuit QED~\cite{circuitBerry}. The JC results agree well with these experimentally extracted phases. One should remember, though, that the Berry phase is a strict adiabatic result, and no experiment can ever reach this limit, mainly due to definite coherence times but also possibly due to finite lifetimes of the experimentalists. In this respect, to be mathematically correct their results are rather measurements of the geometric Aharonov-Anandan phase~\cite{aa}, for which my results do not apply. The natural question then follows; What are the time-scales for the effects of the counter rotating terms to become visible? Numerically answer this question would be very demanding. Almost all numerical methods rely on some sort of truncation, and as the characteristic time-scale grows very large it will most likely imply that the truncation limit has to be pushed further and further in order to keep the numerical accuracy, implying increasing computational power. Related to this issue, we note that there are some recent works contradicting my results~\cite{cont}. Both these use series expansions for approximating the solutions of the Rabi model, and I argue throughout this work that in the adiabatic limit one must be very careful with any neglected small terms, and any truncation of the series expansion must thereby be carefully justified.

\end{document}